\documentclass[a4paper,fleqn,usenatbib]{mnras}

\usepackage{newtxtext,newtxmath}
\usepackage{times}
\usepackage[T1]{fontenc}
\usepackage{ae,aecompl}
\usepackage{supertabular}
\usepackage{longtable}

\usepackage[dvipsnames]{xcolor}

\usepackage{graphicx}	
\usepackage{amsmath}	
\usepackage{amssymb}
\usepackage{float}
\newcommand{\overbar}[1]{\mkern 1.5mu\overline{\mkern-1.5mu#1\mkern-1.5mu}\mkern 1.5mu}

\title[Distances and the role of binarity]{Unbiased TGAS$\times$LAMOST distances and the role of binarity}
\author[J. Coronado et al.]{Johanna Coronado,$^{1}$\thanks{E-mail: coronado@mpia.de} Hans-Walter Rix$^{1}$ and Wilma H. Trick$^{1,2}$
\\
$^{1}$Max-Planck-Insitut f\"ur Astronomie, K\"oningstuhl 17, D-69117 Heidelberg, Germany\\
$^{2}$Max-Planck-Insitut f\"ur Astrophysik, Karl-Schwarzschild-Str. 1, D-85748 Garching b. M\"unchen, Germany
}
\date{Accepted XXX. Received YYY; in original form ZZZ}

\pubyear{2018}

\begin{document}
\label{firstpage}
\pagerange{\pageref{firstpage}--\pageref{lastpage}}
\maketitle

\begin{abstract}
Spectrophotometric distances to stars observed by large spectroscopic surveys offer a crucial complement to parallax distances that remain very important also after the future Gaia data releases.
Here we present a probabilistic approach to modeling spectroscopic information for a subset of 4,000 main sequence stars with good parallaxes ($\sigma_\varpi/\varpi<0.1$) from the LAMOST $\times$ TGAS $\times$ 2MASS cross-match, yielding a precise spectroscopic 
distance estimator with uncertainties of $\sim$6\% for single stars. Unlike previous approaches to this problem, we explicitly account for the individual parallax uncertainties in the model building and fully incorporate the fraction of near-equal binaries of main sequence stars, which would lead to biased distance estimates if neglected. 
Using this model, we estimate the distance for all
(150,000) main sequence stars from LAMOST Data Release 5, without parallax information. As an application, we compute their orbital actions, where our more precise distances result in 5 times smaller
action uncertainties. This illustrates how future studies of the Milky Way's orbital structure can benefit from using our model. For the fainter and more distant stars of most current spectroscopic surveys, an approach such as the one presented in this work will deliver better distances than Gaia Data Release 2.  
\end{abstract}




\section{Introduction}
In the age of Gaia we will have access to $\sim 10^{9}$ stars with some form of parallax and proper motions estimates \citep{gaia1,gaia2}. Combined with information from spectroscopic surveys, the full 6D stellar position-velocity phase-space measurements will allow us to study the dynamics and evolution of our Galaxy. For stars with spectra, the distance estimates will often be the dominant source of uncertainty. This is especially true if the distance information comes solely from parallaxes, as those provide only poor distance estimates for the majority of sources in the Gaia catalog \citep{BailerJones2015}.
Fundamentally, stellar distances can be derived either from direct parallax estimates, or from the extinction-corrected flux if the intrinsic luminosities can be inferred from independent astrophysical information. 
Even after the advent of Gaia Data Release (DR) 2, distance estimates beyond parallaxes will be crucial. Luminosities can be inferred from spectral parameters \citep[$\log g$, T$_{\rm eff}$, etc.;][]{queiroz}, from objects classification \citep[e.g. RR~Lyrae;][]{Sesar2017}, or from asteroseismology \citep{Rodrigues}.

 In general, useful distance constraints for stars come from both parallaxes and spectra. For example, the Tycho Gaia Astrometric Solution (TGAS) in Gaia DR1 has already provided parallaxes for 2.5 million stars  in the solar vicinity \citep[d$\lesssim$ 200 pc;][]{TGAS1,TGAS2}, but they are not precise enough to be used individually and to probe much larger distances \citep{queiroz}. 

Several ground-based dedicated spectroscopic surveys targeting individual stars have become available in the last few years: the Sloan Extension for Galactic Understanding and Exploration \citep[SEGUE;][]{yanny}, the Apache Point Observatory Galactic Evolution Experiment (APOGEE, \citet{apogee2}), the RAdial Velocity Experiment \citep[RAVE;][]{rave}, the Gaia-ESO \citep[GES;][]{gaia-eso}, Galactic Archaeology with HERMES \citep[GALAH;][]{galah,sven}, the Large sky Area Multi-Object fiber Spectroscopic Telescope \citep[LAMOST;][]{LAMOST1,LAMOST2}, the Experiment for Galactic Undertanding and Exploration \citep[LEGUE;][]{legue}, among others. These have provided valuable data that will allow us to comprehensively study the chemical composition and structure of our Galaxy.

This calls for methods to determine optimal distance estimates that incorporate both parallaxes and spectral information. So far, several works have focused on this goal by making use of the Red Clump as a standard candle \citep{hawkins,ruiz}, by making use of all the stars common in TGAS and RAVE \citep{mcmillan}, or by using several spectroscopic surveys, such as APOGEE, RAVE, GES and GALAH \citep{queiroz}. However, none of these works considered unresolved binary stars in their modeling for estimating the distances. In this work we follow a similar theoretical approach as the aforementioned authors, but incorporate binarity explicitly, which becomes important for main sequence stars.
We illustrate our approach in a study that combines LAMOST data with TGAS. 

This is not the first effort to combine these two surveys. \citet{schonrich} already worked towards assessing distances in the TGAS $\times$ LAMOST cross-match, but only using parallax information. In a different study, \citet{LAMOST3} estimated the absolute magnitude directly from LAMOST spectra, obtaining the distance moduli for 50,000 stars with a TGAS-based magnitude error smaller than 0.2 mag, with a 12 percent error in distance. However, as the authors discuss, these results are obtained with very high signal-to-noise ratio spectra with a median value of 150, which will not be available for the majority of stars in large surveys such as Gaia.

Here, we build a probabilistic model that combines parallax and spectroscopic information, using a subset of stars with precise parallaxes ($\sigma_{\varpi}/\varpi$ <0.1) to build a model for their mean absolute magnitude. This model is then applied to the entire sample of main sequence stars from LAMOST DR5.

\begin{figure*}
\centering
	\includegraphics[width=\textwidth]{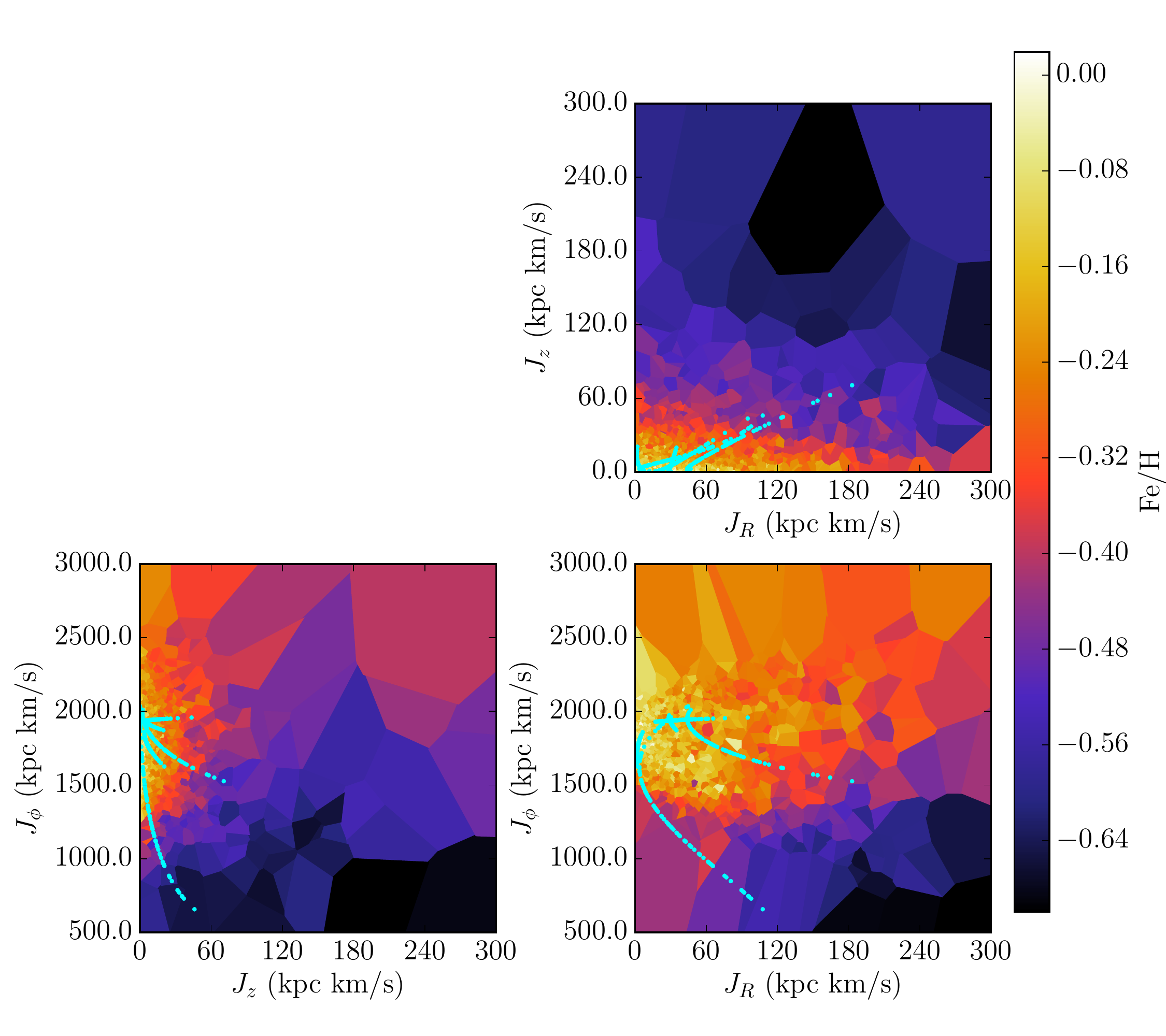}
    \caption{Action distribution ($J_{\phi}$, $J_{R}$, $J_{z}$) for $\sim$ 150,000 main sequence stars of the LAMOST sample with proper motions from the GPS1 catalog \citep{gps1} color coded by metallicity. Each cell in this voronoi plot contains 100 stars. Overplotted are the Monte Carlo sampled error ellipses of 5 example stars (in cyan) that result from transforming the measurement uncertainties on the TGAS parallax into action space. This illustrates both the complexity of the stellar disc in action-abundance space, as well as the need for more precise distance estimations. Here, we give just a very short overview of the rich structure in actions and metallicity, of which a detailed description is beyond the scope of this work: Most stars in the Galactic disc are on near circular orbits $(J_R\sim0,J_z\sim0)$. The overdensity of stars at $J_\phi = R \times v_T \sim 8~\mathrm{kpc} \times 220~\mathrm{km/s} = 1760~\text{kpc km/s}$ is due to the LAMOST survey volume being confined to the solar neighbourhood around $R_\odot\sim8~\text{kpc}$ (see also Fig.~\ref{fig:x_y_x}). In the vertical action $J_z$ we see the well-known vertical metallicity gradient in the disc (e.g. \citealt{2008ApJ...684..287I}). 
The low metallicities at $J_\phi < 1300~\text{kpc km/s}$ are a selection effect of the LAMOST survey, which preferentially selects high-$z$, low-[Fe/H] stars at smaller radii (see Fig.~\ref{fig:x_y_x}). At large $J_z$ the apparent metallicity gradient, rising with increasing $J_\phi$, can be traced back to the $v_T$-vs.-[Fe/H] relation of the thick disc (see e.g. \citealt{2013AA...560A.109H}).} 
    \label{fig:actions}
\end{figure*}

One of the direct applications of precise distances is to improve the determination of stellar orbits in the Milky Way, e.g. characterized by their action distribution ($J_{\phi}$, $J_{R}$, $J_{z}$), see Sec.~\ref{sec:orbital_actions} for an introduction to orbital actions. 
In many circumstances it will remain the case that for stars with spectra, the distance uncertainties --  based on parallaxes alone -- will dominate the uncertainties in calculating the orbits or actions. In Fig.~\ref{fig:actions} we show the distribution of $\sim150,000$ main sequence stars from the LAMOSTxSDSS/GPS1 cross-match in action space, color-coded by the average metallicity per Voronoi bin. This illustrates the richness of structure in the Galactic disc in terms of stellar orbits and chemical abundances. To characterize the complexity of the stellar disc (e.g. \citealt{2016ApJ...823...30B,2015MNRAS.449.3479S}) and to explain it in the context of Galaxy formation and evolution (e.g. \citealt{2017ApJ...834...27M,2018MNRAS.474.3629G}) has been and will be the objective of many studies. Precise estimation of orbits and actions are particularly crucial in action-based dynamical modeling approaches of the Milky Way (e.g. \citealt{2013ApJ...779..115B,2014MNRAS.445.3133P,2016ApJ...830...97T}), and for studies investigating orbital properties (e.g. \citealt{2018arXiv180406379W}) or integrating stellar orbits (e.g. \citealt{2018arXiv180405894S}), to just name some very recent efforts. Fig.~\ref{fig:actions} also shows how measurement uncertainties in parallaxes from TGAS translate into widespread uncertainties in action space (see Sec.~\ref{sec:orbital_actions}). In this work we will, drawing on our model for main sequence absolute magnitudes, illustrate the improvement of orbits with better distances. 

The structure of the paper is as follows: In Sec.~\ref{sec:data} we describe the data used, in Sec.~\ref{sec:model} we present our probabilistic model, while in Sec.~\ref{sec:distances} we show the results of the best fit parameters of this model, and show the effects on the computation of orbital actions using the new estimated distances. We finally summarize and discuss the possible implications of our results in Sections~\ref{sec:discussion} and~\ref{sec:summary}.

\section{Data description}
\label{sec:data}
\begin{figure}
\centering
	\includegraphics[width=\columnwidth]{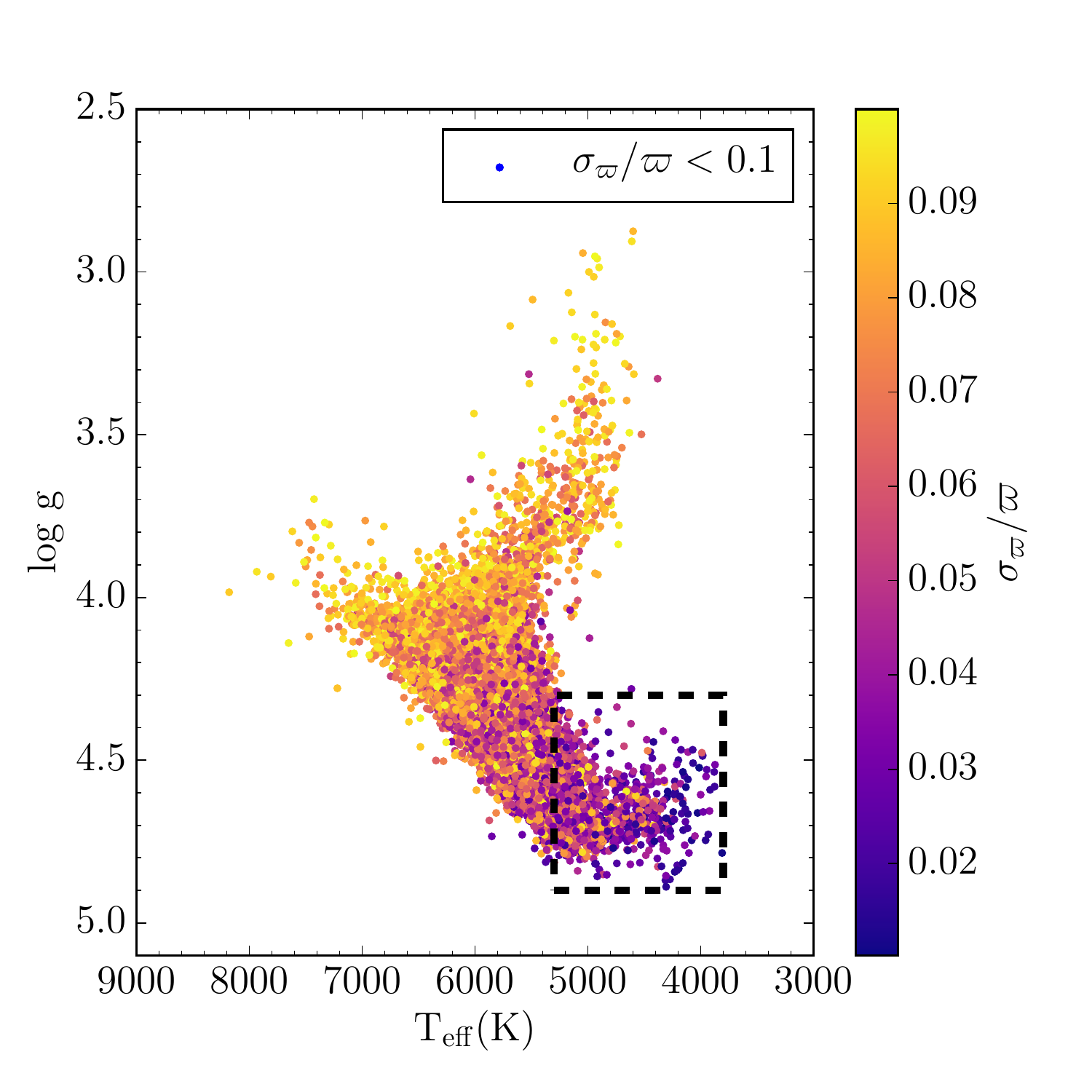}
    \caption{Distribution in the Kiel diagram of the L$\otimes$T~10\% sample. We observe that the coolest stars are most nearby and have the best (fractional) parallaxes. The black dashed rectangle highlights main sequence stars with 3800 K <$T_{\rm eff}$ <5300 K and $\log$ g > 4.2.}
    \label{fig:LAMOST_parallaxes}
\end{figure}

\begin{figure}
\centering
	\includegraphics[width=\columnwidth]{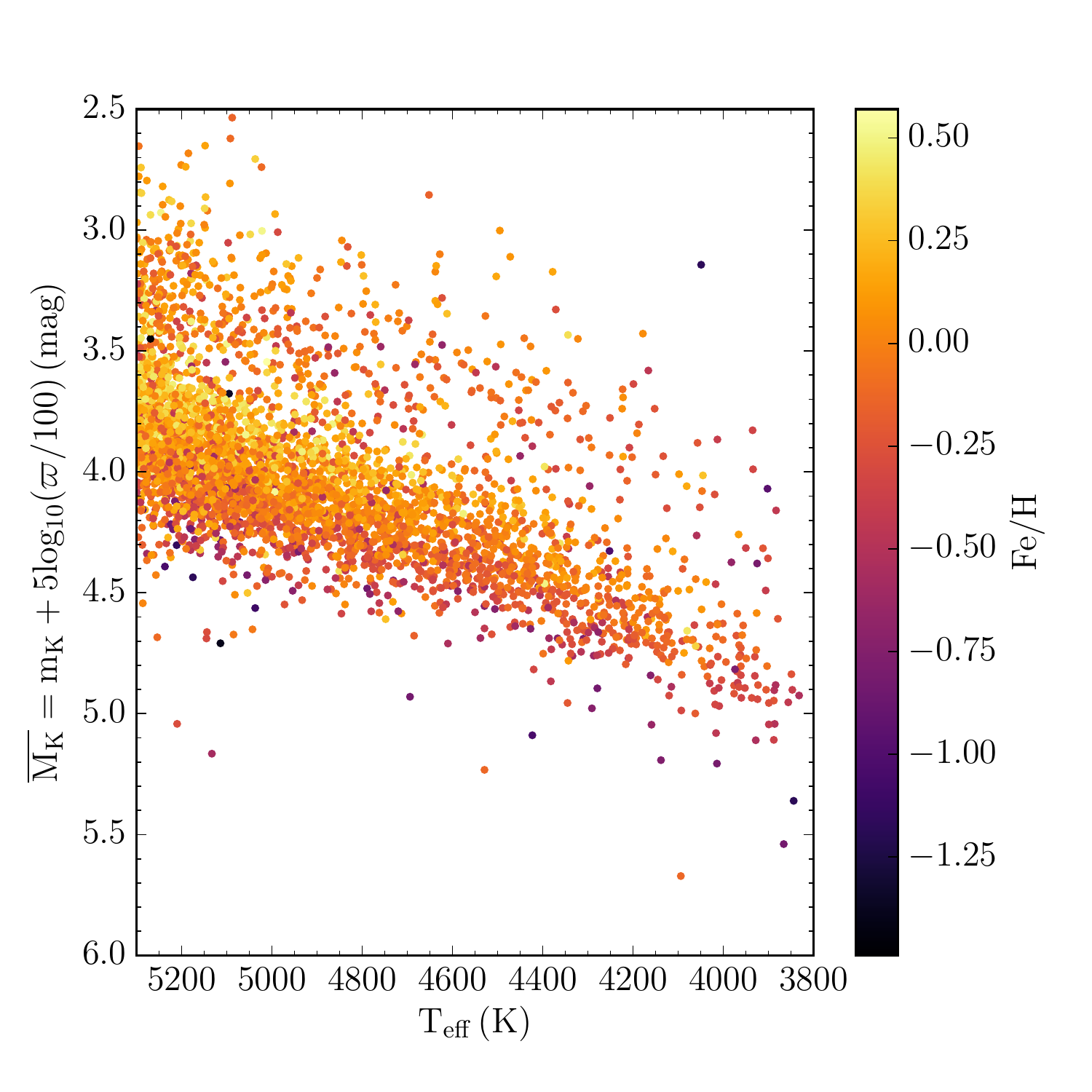}
    \caption{The mean absolute magnitude $\overbar{M_{K}}$ of MS stars in the L$\otimes$T~10\% sample plotted against $\mathrm{T_{eff}}$ and color coded by metallicity. The sequence of presumed binary stars appears shifted roughly 0.6 mag towards brighter magnitudes from the primary sequence located at $\overbar{M_{K}}\sim 4$ for $T_{\rm eff}=5000$~K. The units of $\varpi$ are in milliarcseconds in this figure.}
    \label{fig:mk_parallax}
\end{figure}
In this section we describe how we construct the sample that we use to build our model for the mean absolute magnitude, that we then adopt to estimate the distances. First, we obtain the spectroscopic parameters $(T_\text{eff}, \log g, \text{[Fe/H]})$ from LAMOST DR5 \citep{2011RAA....11..924W,2014IAUS..306..340W}, and obtain their K band magnitude from a cross-match with \textit{2MASS}. Then we cross-match this sample with TGAS, which results in $\sim$ 150,000 stars; hereafter, we will refer to this sample as L$\otimes$T. From L$\otimes$T 40,000 stars have ``good'' parallaxes with $\sigma_\varpi/\varpi < 0.1$. We will refer to this subsample as L$\otimes$T~10\%, and we show its distribution in the Kiel diagram in Fig.~\ref{fig:LAMOST_parallaxes}. The remainder of L$\otimes$T contains stars with mostly poor parallaxes, hence it is of paramount importance to develop a method to obtain distances for those stars beyond parallax information alone.

From the L$\otimes$T~10\% sample, we consider main sequence (MS) stars with 3800~K<$T_{\rm eff}$<5300~K and $\log g$>4.2. This sample is highlighted in Fig.~\ref{fig:LAMOST_parallaxes} with a dashed black rectangle. In Fig.~\ref{fig:mk_parallax} we show this subset of $\sim$4,000 stars where we observe two sequences in the $\overbar{M_{K}}$ vs. $T_\text{eff}$ plane. The primary sequence of presumed single stars is located at $\overbar{M_{K}}\sim 4$ for $T_{\rm eff}=5000$~K; while the secondary and less prominent sequence is shifted by about $\sim$ 0.6 mag towards brighter magnitudes, presumably reflecting unresolved binaries of comparable brightness. Based on this subset of MS stars in the L$\otimes$T~10\% sample, we proceed to build our model. 
\section{A model for main sequence and binary stars}
\label{sec:model}

In this section we will describe how we obtain stellar distances by modeling the mean absolute magnitude of main sequence stars. We start by modeling the K-band absolute magnitude of main sequence stars 
(of given $\log g$, $T_\text{eff}$ and [Fe/H]) as a sum of two Gaussians, the first one centered at a mean absolute magnitude $\overbar{M_K}$ with dispersion $\sigma_1$ and the second one representing the binary sequence and shifted towards brighter magnitudes $\overbar{M_K}-0.6$ with dispersion $\sigma_2$. 

We follow a Bayesian approach to calculate the probability that our proposed model is associated with the observed data. Then, we can write the probability of the model given the data as:
\begin{equation}
\mbox{p(model|data)} = \frac{\mbox{p(data|model)p(model)}}{\mbox{p(data)}}.
\end{equation}

Here, we will explicitly include the individual true distances to the stars as additional free model parameters to be optimized. 

\subsection{Building a model for $\overbar{M_{K}}$}

We start by considering that the mean absolute magnitude $\overbar{M_{K}}$ of MS stars does not only depend on the effective temperature, $\mathrm{T_{eff}}$, but also on metallicity, [Fe/H], as illustrated in Fig.~\ref{fig:mk_parallax}. Hence, we want to determine the mean absolute magnitude using this spectral information. We propose to directly incorporate the spectroscopic parameters ($T_\text{eff}$, \text{[Fe/H]}, $\log g$) and not rely on colors. This will allow us to determine the scatter in the mean absolute magnitude $\sigma$ directly from the data, as opposed to other works that rely on isochrones, a technique first introduced by \citet{burnett} and used subsequently by several other works, including \citet{carlin} who implemented it with LAMOST data, but did not provide the obtained distances. 

We define the mean absolute magnitude of main sequence stars to be a function of the spectroscopic parameters and expand it up to first order in $\log g, \text{[Fe/H]}$ and second order in $T_\text{eff}$ normalizing each parameter by the mean value of the sample: $\overbar{T_\mathrm{eff}}$ = 4900 K, $\overbar{\text{[Fe/H]}}$= -0.019 and $\overbar{\log g}$ = 4.64. This can be expressed as

\begin{equation}
\begin{aligned}
\overbar{M_{K}}(T_\text{eff},\log g,\text{[Fe/H]} \mid \theta_K) = M_{0} + \mathrm{a_{T}}\frac{T_\mathrm{eff} - \overbar{T_\mathrm{eff}}}{\overbar{T_\mathrm{eff}}} \\
+ \mathrm{a_{T_{2}}}\left(\frac{T_\mathrm{eff} - \overbar{T_\mathrm{eff}}}{\overbar{T_\mathrm{eff}}}\right)^{2} + \mathrm{a_{\mathrm{logg}}}(\log g - \overbar{\log g})\\
+ \mathrm{a_{FeH}}\,(\mathrm{[Fe/H]} - \overbar{\mathrm{[Fe/H]}}).
\label{eq:M_K}
\end{aligned}
\end{equation}

\subsection{The probability function including the binary sequence}

An unresolved binary system of two identical stars has the same color but twice the luminosity of an equivalent single star. Such systems would form a second sequence or ``ridge'' in the Color Magnitude Diagram (CMD) \citep{hurley}, running almost parallel to the main sequence $\sim$ 0.7 mag brighter. 
\citet{elbadry} illustrated that the $T_{\rm eff}$-luminosity tracks of unresolved binary stars with different mass rations $q$ run nearly parallel to the single-star main sequence for $0.8 < q < 1$.
As a good fraction of binaries have mass ratios in this range, such a binary ``sequence''\footnote{This sequence is more precisely a caustic of the binaries' track in the $T_{\rm eff}$-luminosity plane
\citep{elbadry}.} should be a generic feature. Indeed, Fig.~\ref{fig:mk_parallax} shows 
such a binary sequence among field stars (systems with $0.8 < q < 1$), this is what we observe and 
what we will model in this work.

We model the distribution of stars in the predicted mean absolute magnitude $\overbar{M_{K}}$ mentioned in Eq.~\eqref{eq:M_K} as the sum of two Gaussians:
\begin{equation}
p(\overbar{M_{K}}|\theta_{M}) = (1- \mathrm{f_{eqb}})\cdot \mathcal{N}(\overbar{M_{K}},\sigma_{1})\,+\, \mathrm{f_{eqb}}\cdot \mathcal{N}(\overbar{M_{K}}-0.6,\sigma_{2}). \label{eq:absolute_magnitude_distribution}
\end{equation}
We have added the second Gaussian term that accounts for the fraction of near-equal binaries
($0.8 < q < 1$), defined as $\mathrm{f_{eqb}}$. Hereafter, we will use the term binaries to refer to systems in this $q$-range. We denote the joint posterior probability distribution of the proposed model given the observations as $\mathrm{p}(\theta_{M}, \{d_{i}\}|\{\mathrm{D}_{i}\})$. We define the model parameters as
\begin{equation}
\theta_{M} = \{M_{0}, \sigma_{1}, \sigma_{2}, \mathrm{f_{eqb}}, \mathrm{a_{T}},\mathrm{a_{T_{2}}}, \mathrm{a_{log g}}, \mathrm{a_{FeH}}\},
\end{equation}
and the data as
\begin{equation}
\{\mathrm{D}_{i}\}$ = $\{\varpi_{i}, \sigma_{\varpi_{i}}, m_{i}, \sigma_ {m_{i}}, T_{\mathrm{eff},i}, \log g_{i}, \mathrm{[Fe/H]}_{i}\}. 
\end{equation}
We denote the distances to each star as $d_{i}$. Using Bayes theorem we can write the posterior probability of our proposed model as: 
\begin{equation}
\mathrm{p}(\theta_{M},\{d_{i}\}|\{\mathrm{D}_{i}\}) \propto \mathrm{p}(\theta_{M})\prod_{i}\mathrm{p}(\mathrm{D}_{i}|\theta_{M},d_{i})\,\mathrm{p}(d_{i}|\theta_{M}), \label{eq:posterior_probability}
\end{equation}
Where $\mathrm{p}(\theta_{M})$ are the model priors, $\mathrm{p}(\mathrm{D}_{i}|\theta_{M},d_{i})$ is the likelihood function and $\mathrm{p}(d_{i}|\theta_{M})$ is the prior distance for each star. We use the exponentially decreasing volume density prior from \citet{bailer-jones1} for the distances, while we consider flat priors for the rest of the model parameters.

\subsection{A note on extinction}
\label{sec:note_ext}
In principle, we would need to correct the apparent magnitudes for dust extinction before applying our model. To gauge the importance of extinction for our sample we check the reddening values for each star from the 3-dimensional dust map by \citet{green}. This map provides the best-fit for E(B-V) in each distance slice, for which we use 1/$\varpi$. We note that this is an acceptable approximation given the fact that our sample extends just up to $d\lesssim 200$~pc, and the errors in the parallaxes are small ($\sigma_\varpi/\varpi < 0.1$). We take the extinction coefficient for the K band ($\mathrm{R_{K}}$) from \citep{yuan} to convert from reddening to the K band extinction, as $A_{K} = \mathrm{R_{K}}\,\times\,\mathrm{E(B-V)}$. However, we find that $A_{K} <0.1$ and its mean value is $\approx$ $6x10^{-2}$, therefore reddening is not important in this sub-sample. 
We emphasize that this procedure is done to test that the sample we use in building our model is dust free. But in order to obtain reliable distances using a different sample, a correction for extinction must be performed to the apparent magnitude. 

\section{Spectrophotometric distances}
\label{sec:distances}

\subsection{Finding the best-fit model and distances to stars with good parallaxes} \label{sec:model_with_good_parallaxes}

\subsubsection{Including measurement uncertainties in the likelihood}

We assume that the observed apparent magnitude $m_{i}$ of each star is the outcome from measuring the brightness of a star with true absolute magnitude $M_{Ki}(T_\text{eff},\log g, \text{[Fe/H]} \mid \theta_M)$ at a true distance $d_i$ with some measurement uncertainty $\sigma_{m_i}$. The observed parallax $\varpi_i$ is assumed to be drawn from a Normal distribution described by the true parallax $1/d_i$ and the observational uncertainty $\sigma_{\varpi_i}$. The joint likelihood for each star is therefore:

\begin{equation}
\mathrm{p}(\mathrm{D}_{i}|\theta_{M},d_{i}) = \mathrm{p}(\varpi_{i},\sigma_{\varpi_{i}}|d_{i})\mathrm{p}(m_{i}, \sigma_{m_i}|\theta_{M},d_{i}), \label{eq:measurement_uncertainties_in_likelihood}
\end{equation}
where $\mathrm{p}(\varpi_{i},\sigma_{\varpi_{i}}|d_{i}) = \mathcal{N}(\varpi_{i}|1/d_{i},\sigma_{\varpi_{i}})$ is the parallax likelihood, defined as a Gaussian evaluated at $\varpi_{i}$ and centered around a mean $1/d_{i}$ with dispersion $\sigma_{\varpi_{i}}$. We proceed analogously with the apparent magnitude: $\mathrm{p}(m_{i}, \sigma_{m_{i}}|\theta_{M},d_i) = \mathcal{N}(m_{i}|m_{i,\text{true}},\sigma_{m_{i}})$, where $m_{i,\text{true}}$ is the predicted apparent magnitude,

\begin{equation}
 m_{i,\text{true}} = \overbar{M_{Ki}} +A_K + 5\mathrm{log}_{10}(d_{i}) - 5,
\end{equation}
and $\overbar{M_{Ki}}$ has a possible range of values predicted by the distribution in Equation \eqref{eq:absolute_magnitude_distribution} given the spectroscopic parameters of the $i$-th star and the model parameters $\theta_M$. The extinction term $A_K$ is only included if necessary (see discussion in Sections~\ref{sec:note_ext}~and~\ref{sec:ext}). The likelihood function in Equation \eqref{eq:measurement_uncertainties_in_likelihood} is used in the posterior probability distribution in Equation \eqref{eq:posterior_probability}. 

\begin{figure}
\centering

\includegraphics[width=\columnwidth]{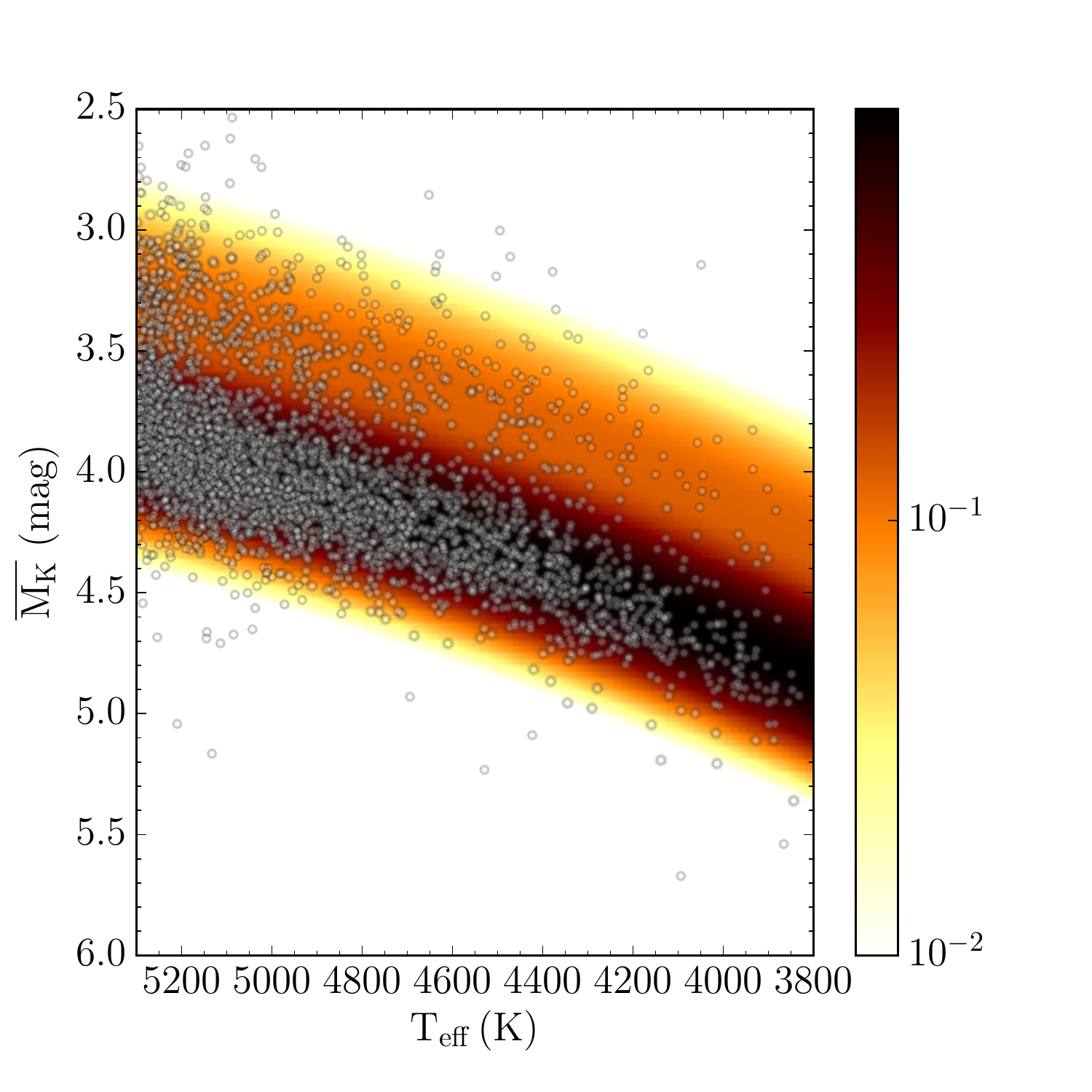}
    \caption{Mean absolute magnitude model fit to MS stars in the L$\otimes$T~10\% sample. Importantly, the model incorporates the binary sequence. The color represents the density of the model {\it pdf} for the mean absolute magnitude. This is the best-fit model (parameters from Table~\ref{table1}), convoluted with a Gaussian of 0.15 mag reflecting the typical parallax uncertainty, for direct comparison with the data.}
    \label{fig:mk_model}
\end{figure}

\begin{figure}
\centering
\includegraphics[width=\columnwidth]{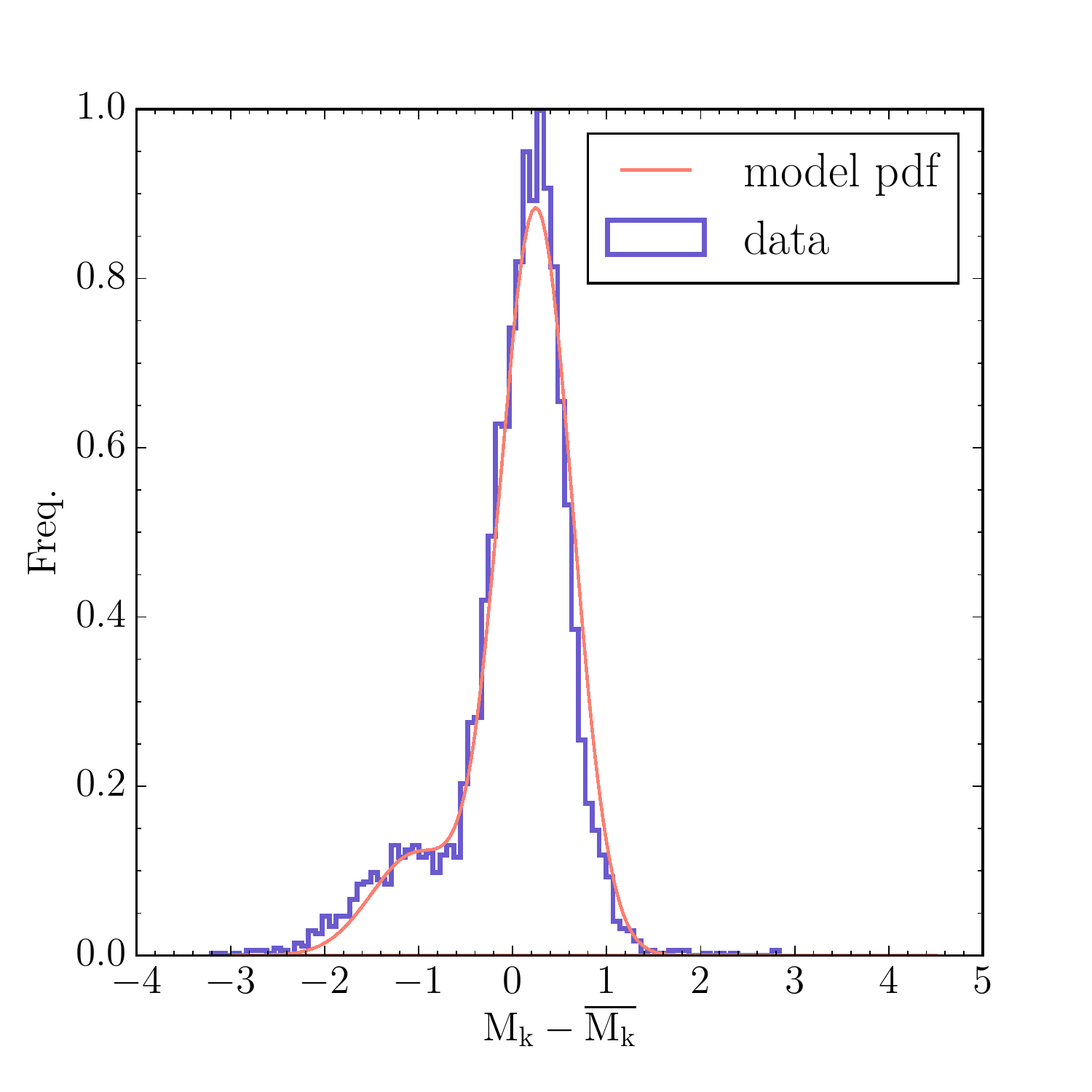}
\caption{Normalized number of stars in red and probability for $M_{K}$ (at a fixed $T_\text{eff}$) minus the mean absolute magnitude $\overbar{M}_{K}$ at that same $T_\text{eff}$ in blue. The data set and model {\it pdf} correspond to the ones represented in Fig.~\ref{fig:mk_model}.}
    \label{fig:mk_model2}
\end{figure}

\subsubsection{Exploring the space of model parameters}

We use emcee \citep{emcee}, a python implementation of Goodman \& Weare's Affine Invariant Markov chain Monte Carlo (MCMC) Ensemble sampler to draw samples from the posterior distribution in Equation \eqref{eq:posterior_probability} for our model parameters and distances. The parameter space has $(8+N)$ dimensions, the 8 parameters $\theta_M$ and the $N$ distances $d_i$ when fitting $N$ stars. To reduce the dimensionality of this optimization problem, we start by considering a subset of 100 stars in the MS L$\otimes$T~10\% sample from Fig.~\ref{fig:mk_parallax} to establish the parameter values of our model; these stars are randomly drawn. We sample the posterior distribution to find the best parameters that fit our model. In a second step, we use these values as the initial guess for sampling all of the stars within the MS L$\otimes$T~10\% sample but now marginalizing over the distances. At the end of this step we then have obtained the best parameters for the entire MS L$\otimes$T~10\% sample which we illustrate in Fig.~\ref{fig:mk_model}. The best fit model parameters do not depend on the exact choice of 100 stars from the L$\otimes$T~10\% sample.

Fig.~\ref{fig:mk_model} illustrates the {\it pdf} density of the modeled $\overbar{M_{K}}$ defined in Equation~\eqref{eq:M_K}. For this purpose we use the best fit parameters presented in Table~\ref{table1}. Note again that we did not model the number of stars in the ($\overbar{M_{K}}$,$T_\text{eff}$, $\log g$, $\rm{[Fe/H]}$) plane, but rather the value of $\overbar{M_{K}}$ given ($T_\text{eff}$, $\log g$, $\rm{[Fe/H]}$).
In order to correctly represent the model {\it pdf} of the mean absolute magnitude in Fig.~\ref{fig:mk_model} we also have to incorporate the error in the magnitude. We are considering stars with 10\% error in parallax to construct the model which roughly translates into an error of 0.15 mag. Because we are already modeling the mean absolute magnitude as a Gaussian, we incorporate this error in Equation \eqref{eq:absolute_magnitude_distribution} by performing a convolution. In Fig.~\ref{fig:mk_model2} we show a comparison between the data set and the model {\it pdf} of the mean absolute magnitude also shown in Fig.~\ref{fig:mk_model}. 
We observe that our model for the mean absolute magnitude captures the essence of the data, including binarity.

\begin{figure*}
\centering
    \includegraphics[width=\textwidth]{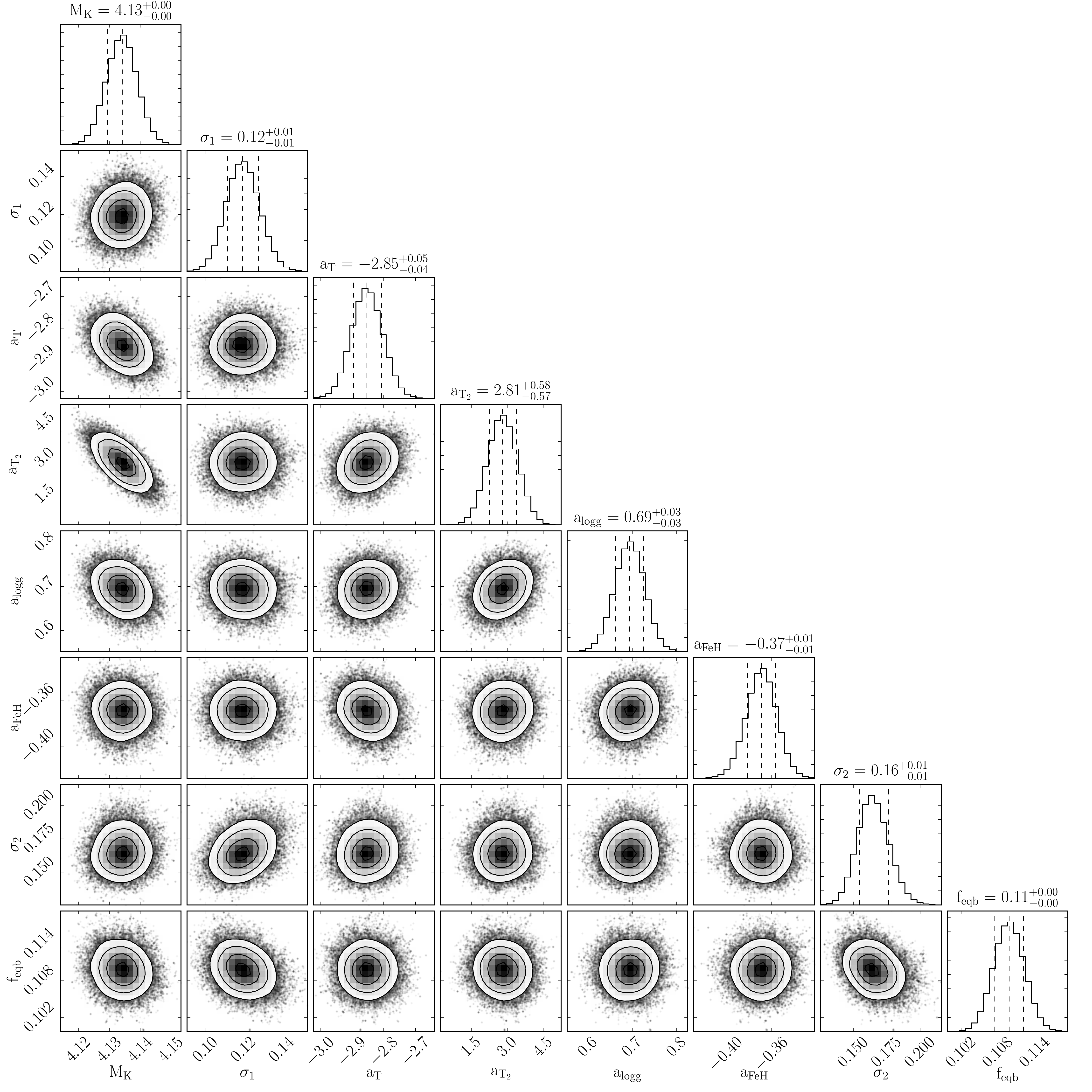}
    \caption{Corner plot showing the samples from the posterior probability for each parameter of our model for the main sequence considering binary stars. For building the model we use stars in the MS L$\otimes$T~10\% sample.}
    \label{fig:corner}
\end{figure*}

In Fig.~\ref{fig:corner} we show the samples from the posterior probability distribution for each of the parameters of our model. All parameters are well constrained with relative uncertainties of between 1\% and 6\%, and only weak covariances. 

Having established the best fit for our model parameters, we can now proceed to obtain the distances for the entire MS L$\otimes$T~10\% sample. 

One advantage of the type of modeling we have applied is that after finding the best fit for our parameters, we can treat the model as fixed and apply it to many more stars that have very bad parallax or no parallax information at all.

\begin{table}
\centering
\caption{Results obtained with emcee for the parameters of our model. In Fig.~\ref{fig:corner} we show the posterior probability for each of these parameters.}
\label{table1}
\begin{tabular}{|l|l|c|}
\hline
model parameter & & best fit\\\hline
peak abs. magnitude in $K$-band & $\mathrm{M_{0}}$ (mag) & $4.134^{+0.004}_{-0.005}$\\
width of abs. magnitude distribution & $\sigma_{1}$ (mag)& $0.119 \pm 0.008$\\
prefactor of $(T_\text{eff}-\overbar{T_\text{eff}})$ term & $\mathrm{a_T}$ & $-2.853^{+0.046}_{-0.043}$\\
prefactor of $(T_\text{eff}-\overbar{T_\text{eff}})^2$ term & $\mathrm{a_{T_{2}}}$ &  $2.809^{+0.583}_{-0.567}$  \\
prefactor of $(\log g - \overbar{\log g})$ term & $\mathrm{a_{logg}}$& $0.694^{+0.031}_{-0.032}$\\
prefactor of $(\text{[Fe/H]}-\overbar{\text{[Fe/H]}})$ term & $\mathrm{a_{FeH}}$& $-0.369 \pm 0.012$ \\
width of binary sequence & $\mathrm{\sigma_{2}}$ (mag)& $0.165^{+0.011}_{-0.010}$ \\
binary fraction (equal mass binaries) & $f_\text{eqb}$ & $0.110 \pm 0.002$ \\ \hline
\end{tabular}
\end{table}

\subsection{Distances to stars with no useful parallaxes}

\subsubsection{Applying the best-fit model for $\overbar{M_{K}}$}

With this model at hand, we can determine distances to entire LAMOST DR5 MS sample, most of which has currently no parallax information. We combine it with the GAIA $\times$ PS1 $\times$ SDSS (GPS1) catalog \citep{gps1}. From this cross-match we obtain $\sim$ 150,000 stars (hereafter, L$\otimes$G) with proper motions from GPS1 and spectroscopic information $\log g$, $T_\text{eff}$, $\rm{[Fe/H]}$ and line-of-sight velocities from LAMOST also apparent magnitudes in the K band from \textit{2MASS}, and in the G band from Gaia. 

With the final parameters shown in Fig.~\ref{fig:corner} we proceed to apply our model to this sample, while the L$\otimes$T~10\% data and model in Sec.~\ref{sec:model} and Sec.~\ref{sec:model_with_good_parallaxes} could be treated as dust-free, this is no longer true for the whole LAMOST sample. So we first correct  the apparent K-band magnitude for extinction using the method described in detail below in Sec.~\ref{sec:ext}. 

After this step we proceed to calculate their corresponding spectrophotometric distances using the parameters obtained from the best-fit model illustrated in Fig.~\ref{fig:corner} and Table~\ref{table1}. 
\newline\newline
The posterior distribution presented in Equation~\eqref{eq:posterior_probability} provides a complete description of the distance. However, we want to obtain a single value of the distance along with its uncertainty. We do this by taking the median value of the distribution as the single value for the distance to each star, and for the uncertainties we consider the 16th and 84th percentile.

\begin{figure*}
\centering
	\includegraphics[width=\textwidth]{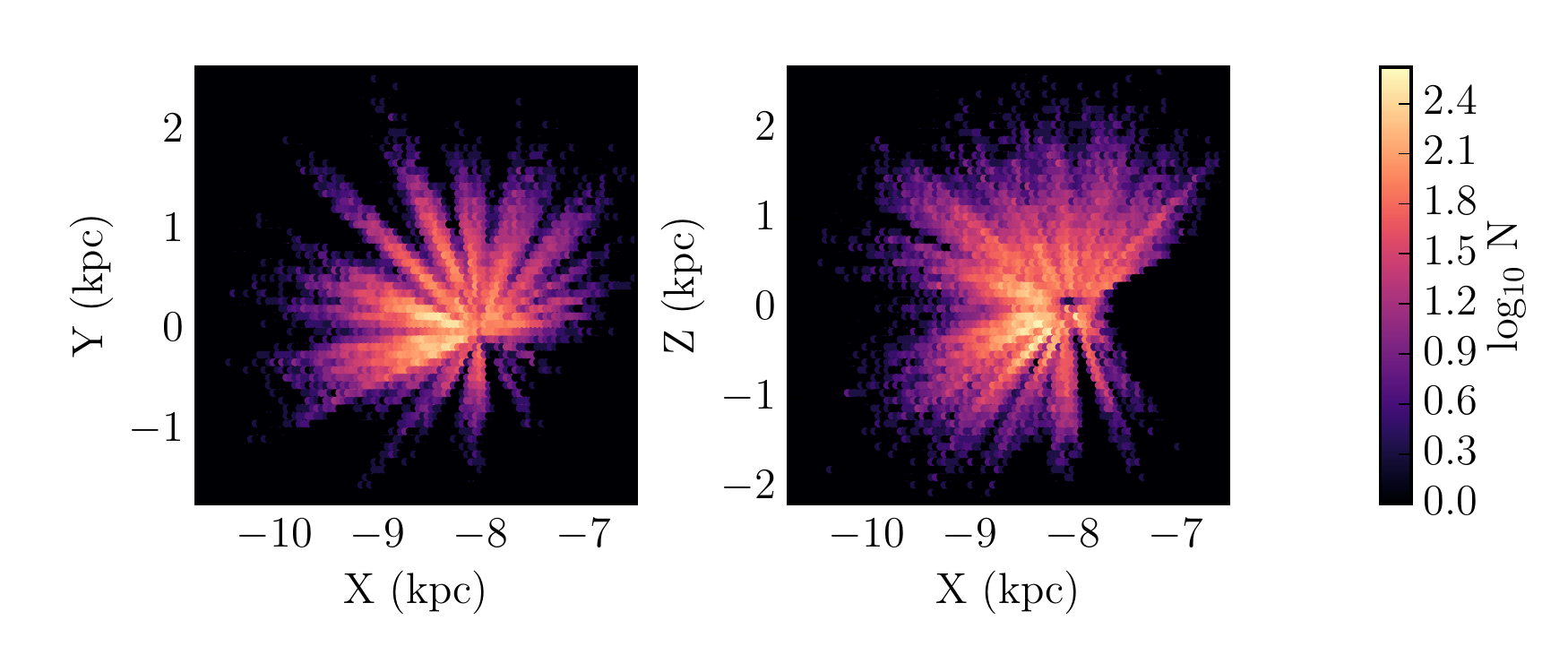}
    \caption{Color coded density distribution in logarithmic bins of the number of stars N in the Galactic X-Y-Z plane for the stars in the L$\otimes$G sample, corrected by extinction and with distances calculated from our model.} 
        \label{fig:x_y_x}
\end{figure*}

With the new calculated distances we illustrate the distribution of these stars in the galactic X-Y-Z plane in Fig.~\ref{fig:x_y_x}, where we assume that $\mathrm{R_{\sun}}$ = 8 kpc and $\mathrm{z_{\sun}}$=0.025 kpc. From these distributions we see that we have a sample more or less confined to the solar neighborhood. 

\subsubsection{Correcting for Extinction}
\label{sec:ext}
\begin{figure}
\centering
	\includegraphics[width=\columnwidth]{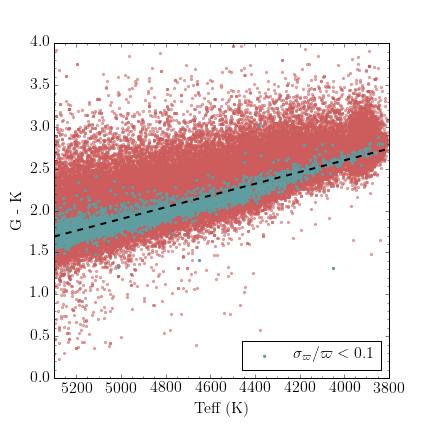}
    \caption{Color G-K vs $T_\mathrm{eff}$ for MS stars including those without parallax information from the L$\otimes$G sample in pink dots. Blue dots show the MS L$\otimes$T~10\% sample, they are nearby and slightly reddened. The dashed line represents the mean location of the blue dots in this diagram and we use it to write an empirical relation to represent the extinction coefficient in the K band.} 
    \label{fig:extinction}
\end{figure}

For the case where we have no parallax information, we cannot retrieve the reddening values from the dust map by \citet{green}. Instead, we use the color G-K as an estimator for reddening, following the relationship between infrared and optical extinction proposed by \citet{cardelli}. We can write the extinction coefficient, $A_{K}$ as a function of color and $T_\text{eff}$ as $A_{K}= f$(G-K|$T_\text{eff}$). Because the value for the $T_\text{eff}$ comes from spectroscopy, it is independent of reddening. In Fig.~\ref{fig:extinction} we plot G-K vs. $T_\text{eff}$ for the MS L$\otimes$T~10\% sample which has good parallax information and the L$\otimes$G sample with no parallax information. We observe that the MS L$\otimes$T~10\% sample is tightly constrained in G-K as a function of $T_\text{eff}$ and therefore it is not strongly affected by dust as we already noted in Sec.~\ref{sec:model}. The dashed line in Fig.~\ref{fig:extinction} represents this empirical relation, and we use it to obtain the extinction coefficients in the K band.

The L$\otimes$G sample has a large spread in G-K, indicating that it is affected by extinction. We quantify the excess in color with respect to the dashed line in Fig.~\ref{fig:extinction} as the amount of extinction.

We proceed to write an empirical relation to obtain $A_{K}$, where we follow \citet{cardelli} and use their eq.1 that expresses the mean extinction law as $<A(\lambda)/A(V)> = a(x) +b(x)/R_{V}$ and Table 3 for the value of a(x). Here, we have ignored the slight color dependence in the transformation between the G and V magnitudes \citep{jordi} and have simply treated the G as V band. We consider that all the points that lay above the dashed line in Fig.~\ref{fig:extinction} are affected by dust, and therefore must be corrected for extinction. The points below are not corrected but still remain in our sample.

We calculate the extinction coefficient $A_{K}$ for each star, and then we can correct the apparent magnitude and therefore proceed to calculate the distances for each star using our model.

\section{Discussion}
\label{sec:discussion}
In the previous sections we have established a model that in the first step relies on a subset of stars with precise parallax measurements. These stars are used to find the best-fit parameters for a mean absolute magnitude model for the MS that depends on spectroscopic information. In a second step, with the established and now fixed model we can obtain spectrophotometric distances. This allows us to obtain improved distances even for the stars that have poor or no parallax information. In the following we further illustrate this.

\subsection{The effect of parallax uncertainty on the recovered distances}
\label{sec:pdf_distance1}
\begin{figure}
\centering
	\includegraphics[width=\columnwidth]{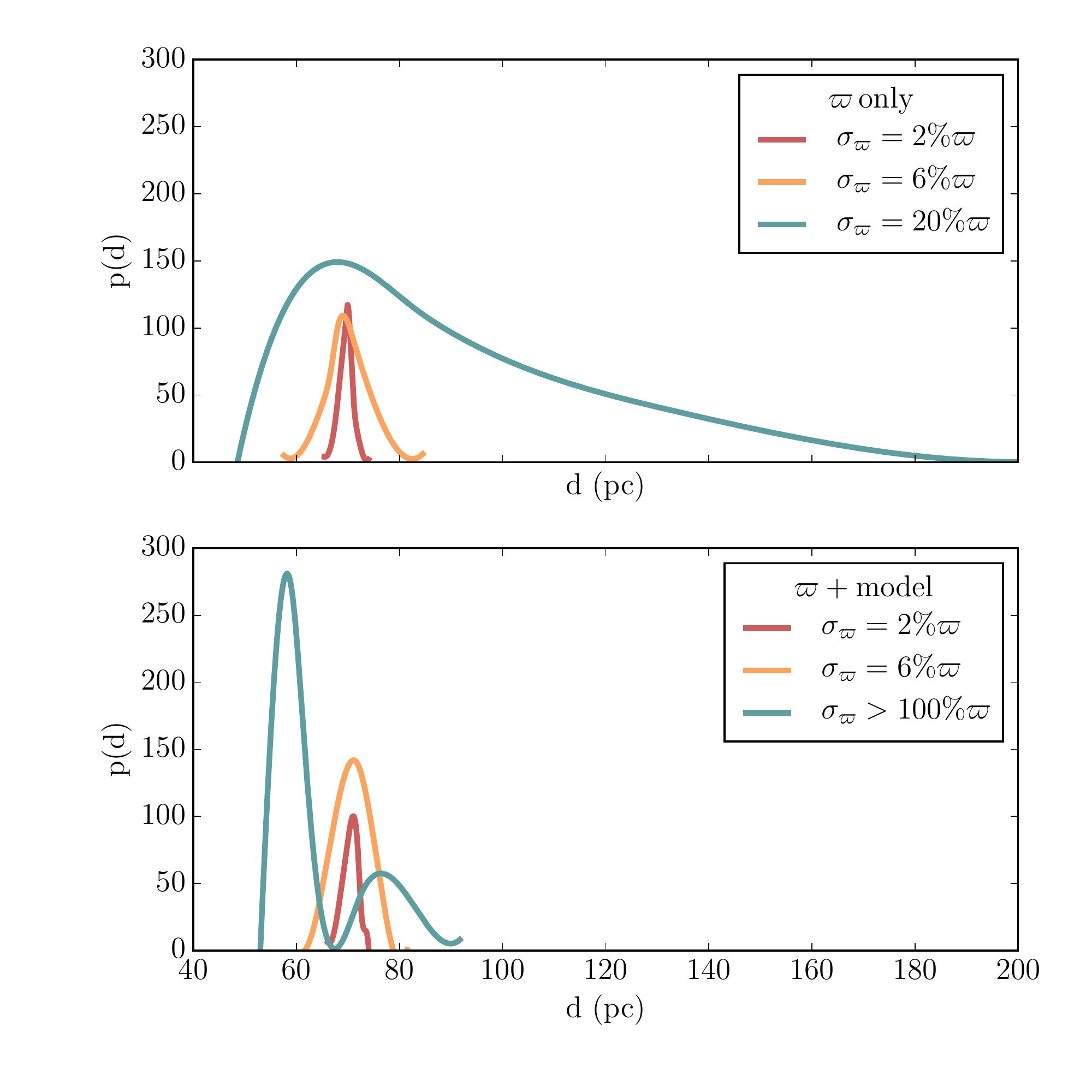}
    \caption{In both panels of this figure we show the result of the posterior distribution of the distance for the same star in different regimes of parallax error. The median value for distance of this star is d =70 pc. The upper panel shows the results considering only $\varpi$ and the lower panel the results when applying our model. We show the results in the regime of a very good parallax with $\sigma_{\varpi}<5\%$ in light red and $\sigma_{\varpi}<10\%$ in yellow. In the bad parallax regime the upper panel shows the result for $\sigma_{\varpi}=20\%$ in light blue and the lower panel for $\sigma_{\varpi}>100\%$.}
   \label{fig:pdf_dists1}
\end{figure}
In this section we will explore how a star's best fit distance changes when we consider different regimes of parallax error. In this experiment, we consider the same star, but vary its $\sigma_\varpi.$ In Fig.~\ref{fig:pdf_dists1} we show the posterior distribution of the distances that we obtain if we apply our model with the best fit parameters from Table~\ref{table1}, for very precise parallaxes and for a very extreme case of poor parallax:

$\sigma_\varpi=2\%, 6\%, >100\% \varpi$. We observe that in the limit of a very good parallax for a star, we obtain a single narrow Gaussian. However, with bad parallax information, we get a bimodal distance distribution, as the model relies entirely on the (also bimodal) absolute magnitude probabilities to obtain the distances. 

Using the same star, we now explore the posterior distribution of the distance that we obtain if we rely only on parallax information. This can be seen in the top panel of Fig.~\ref{fig:pdf_dists1}. Here we also show results for the very good parallax regime, with $\sigma_\varpi = 2\%, 6\%\varpi$.
As an example of the regime of ``poor parallaxes'', we show the results for $\sigma_\varpi = 20\%\varpi$. Even larger parallax uncertainties yield divergent uncertainties on the distance estimates \citep{BailerJones2015}.

\subsection{Estimating orbital actions from data with observational uncertainties}
\label{sec:orbital_actions}
In this subsection we will describe a direct application of our improved distances using the L$\otimes$T sample that contains stars with poor parallax estimates. From the distribution of stellar orbits, we can learn about both the dynamics and formation of the Galaxy. The movement of a star on an orbit can be easily described by the canonical action-angle coordinates (\textbf{$\theta$},\textbf{J}). If we consider an axisymmetric gravitational potential, then these orbits can be fully determined by three integrals of motion \textbf{J} = ($J_{R},J_{\phi},J_{z}$), and they are defined as:
\begin{equation}
J_{i}  = \frac{1}{2\pi}\oint p_{i}dx_{i},
\end{equation}
where the integral is evaluated along the orbit with position $x(t)$ and momentum $p(t)$. The actions \textbf{J} label orbits and each angle variable $\theta$ increases linearly with time and indicate the position of the star along the orbit. $J_{R}$ quantifies the oscillations inwards and outwards in the radial direction, $J_{z}$ quantifies the oscillations in the vertical direction and $J_{\phi}$ is the component of angular momentum. We redirect the reader to \citet{binney2008} Sec. 3.5 for a detailed description of actions.

 \begin{figure}
\includegraphics[width=\columnwidth]{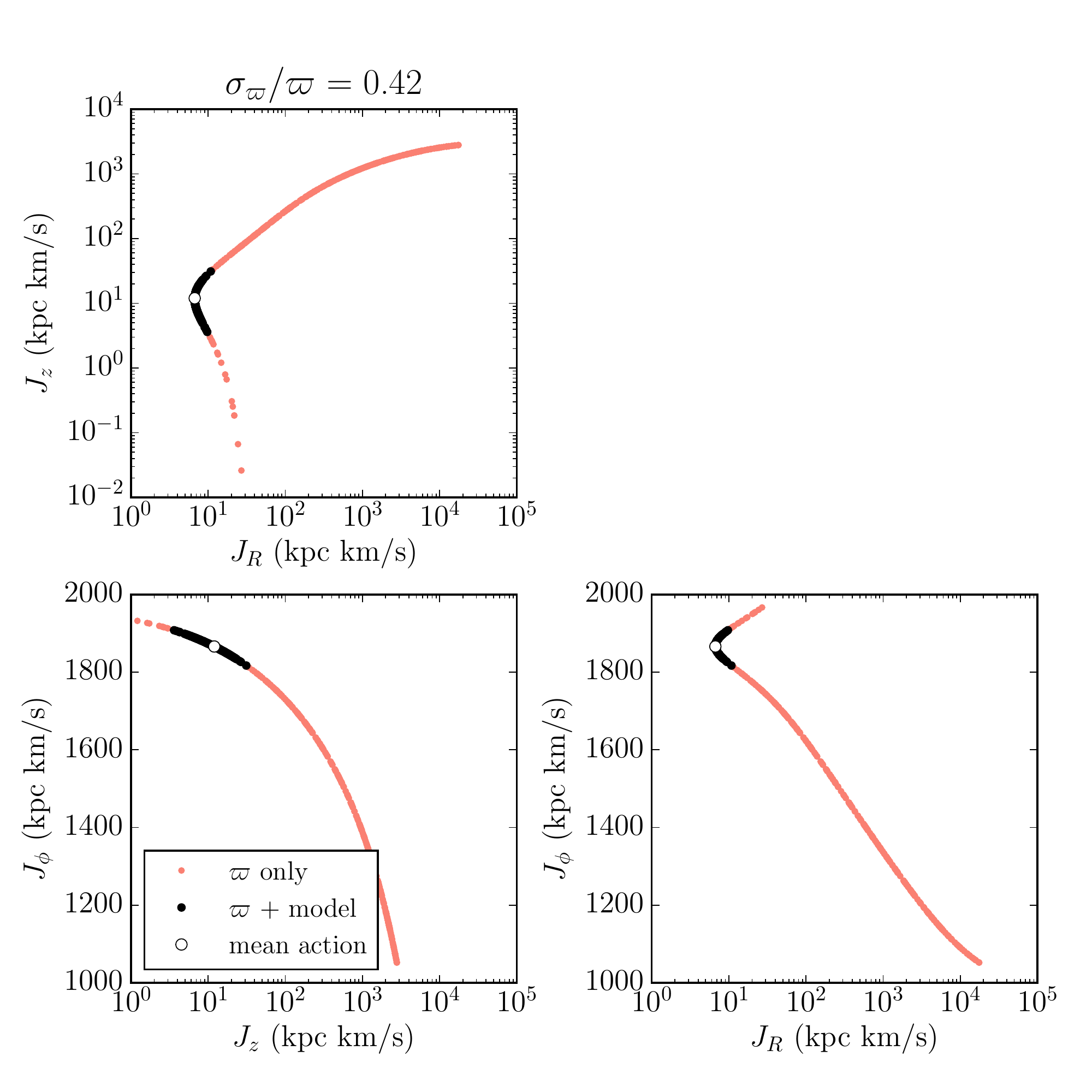}
	\caption{1,000 samples of measurement uncertainty ellipse transformed to action space ($J_{R},J_{\phi}, J_{z}$) performed via Monte Carlo sampling for a star in the bad parallax regime ($\sigma_\varpi = 42\%\varpi$) in the L$\otimes$T sample. These show the extent of the uncertainties in action space when the parallax is very imprecise. We compare the results of our model in black dots and considering only $\varpi$ in red dots. The white dot shows the action's mean measurement.}
    \label{fig:action_samples1}
\end{figure}
\begin{figure}
\centering
	\includegraphics[width=\columnwidth]{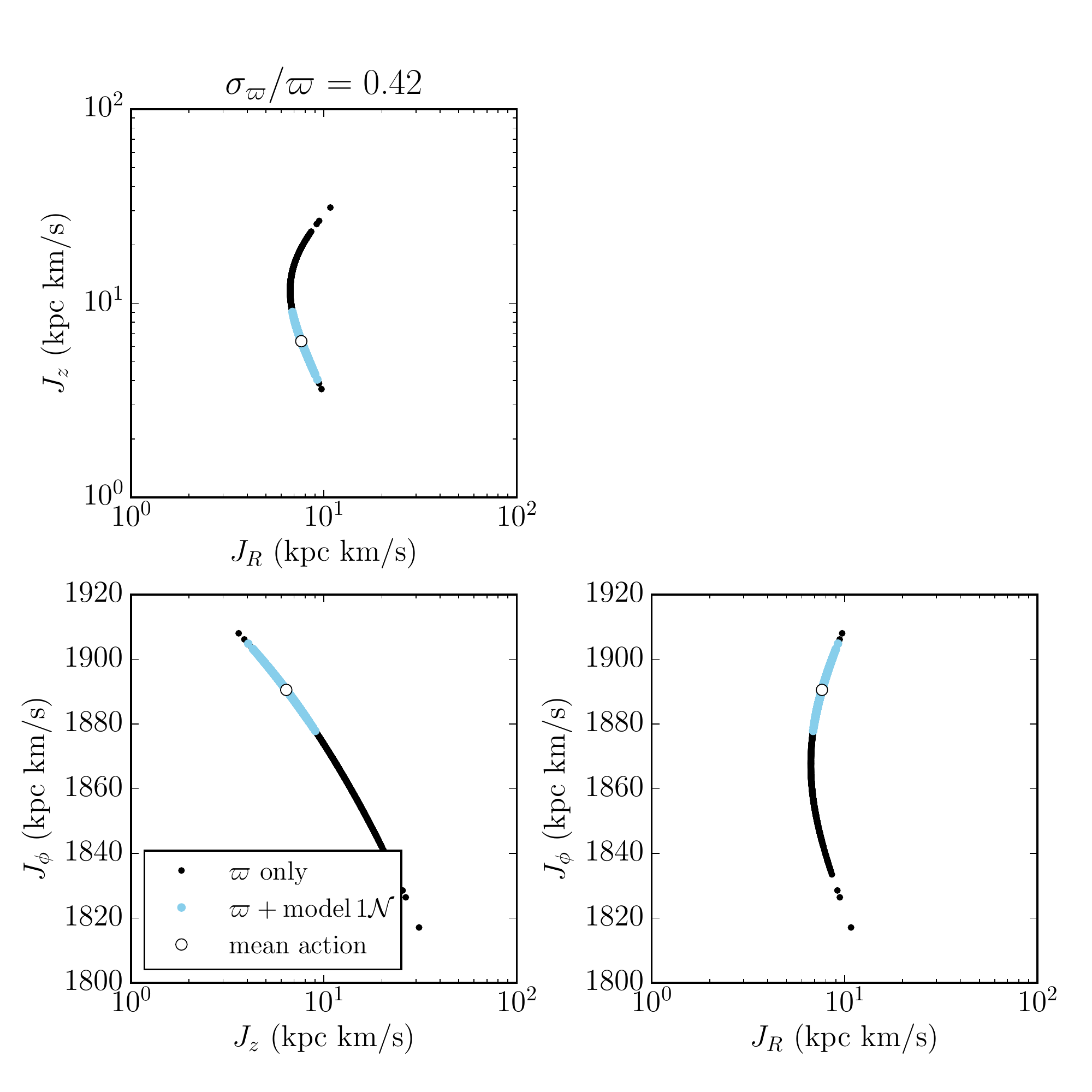}
    \caption{1,000 samples of measurement uncertainty ellipse transformed to action space ($J_{R},J_{\phi}, J_{z}$) performed via Monte Carlo sampling for the same star from Fig.~\ref{fig:action_samples1}. Here we compare the samples resulting from our model that considers binary stars in black dots again and a model that considers one gaussian i.e, only single stars in blue dots. Again, the white dot shows the action's mean measurement.} 
    \label{fig:action_samples2}
\end{figure}

To compute the actions we need the full 6D phase information i.e., velocities ($\mu_{\mathrm{R.A}}$, $\mu_{\mathrm{dec}}$, v$_{los}$) and positions (RA, DEC, distance). For this calculation,
we make use of the \textit{galpy} package, which is a python implementation for galactic-dynamics calculations \citep{bovy}. We consider the simple axisymmetric Milky Way potential with a Miyamoto-Nagai disc, NFW halo and power law bulge that it is implemented in \textit{galpy} as \texttt{MWPotential2014} \citep{bovy}. 
We transform the position and velocities of each star to Galactocentric coordinates, where we consider the position and velocity of the sun for the coordinate transformation to be at (X,Y,Z)= (8, 0, 0.025) kpc and (U,V,W)= (-11 ,230, 7) km/s, respectively. 
From the 6D phase information, the distance is the one with the largest impact on the action distribution uncertainties. We translate the uncertainties from the observations to action space via Monte Carlo sampling of an error ellipse. We convert each sample of the error ellipse from the observable space to Galactocentric cylindrical coordinates and then to actions. We then run 1,000 samples of the error ellipse to explore the extent of distance uncertainties.

\subsubsection{The effect of improved distances}

We compare the distance uncertainties obtained using only parallax information, and our model considering both parallax and spectroscopy. We have shown already in Sec.~\ref{sec:pdf_distance1} by exploring the distance's posterior distribution, that incorporating spectroscopic information becomes especially important in the regime with very bad parallaxes. When exploring the action space this effect is also visible. Relying just on parallaxes causes the measurement uncertainty distribution to spread out over a large portion of action space as can be seen in Fig.~\ref{fig:action_samples1}.

\subsubsection{The effect of binarity}

We can also explore the effect of binarity on the estimate of a stars orbits, seen in action space. We do this by modeling the absolute magnitude with only one Gaussian, not considering the contribution of binary stars. This can be seen in Fig.~\ref{fig:action_samples2}. Here we show the same star as Fig.~\ref{fig:action_samples1}. We note that the model with only just one Gaussian as expected, shows smaller uncertainties in action space. Nevertheless, the effect of considering binary stars does not translate into an important effect in action space.

\begin{figure}
\centering
	\includegraphics[width=\columnwidth]{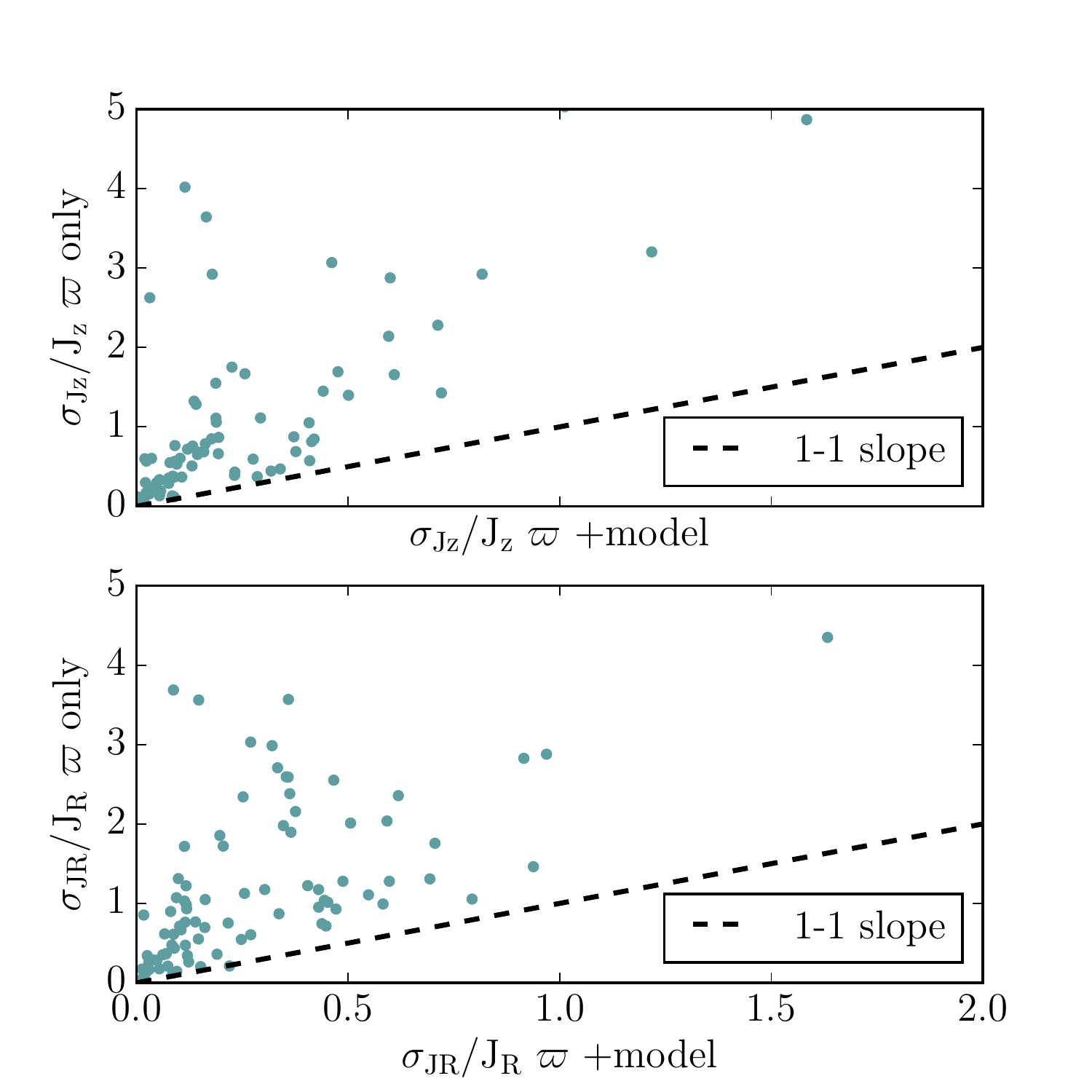}
    \caption{Comparison of the action estimate precision for $\varpi$ only and the model + $\varpi$. We show how our model reduces the uncertainties also in action space by plotting $\sigma_{J}/J$ for stars with $\sigma_{\varpi}/\varpi$ >0.15. We plot $\sigma_{J_{z}}/J_{z}$ in the upper panel and $\sigma_{J_{R}}/J_{R}$ in the lower panel. We observe that $\sigma_{J}/J$ is always smaller when we use our model.} 
    \label{fig:samples_comparison}
\end{figure}

We have shown that having large uncertainties in the distances translates also into large uncertainties in action space, especially when we consider only parallax information, as observed in Fig.~\ref{fig:action_samples1}. This effect becomes even more important when we consider stars with large uncertainties in parallax. But to quantify how much our model actually improves the action space in  Fig.~\ref{fig:samples_comparison} we plot $\sigma_{J}/J$ for the vertical and radial action. To obtain $\sigma_{J}$ we average over the 1,000 action samples per star. In this figure we plot stars with large uncertainties in parallax, $\sigma_{\varpi}/\varpi$ > 0.15. We observe that in all of the cases, $\sigma_{J}/J$ is smaller when we use our model to calculate the distances. Therefore, having precise distances clearly has an impact in the calculation of actions.

\subsection{Spectophotometric distances and DR2}

Now we want to explore in which regime the spectrophotometric distances will be more precise than the parallaxes in the second data release (DR2) of Gaia. We consider the sample L$\otimes$G which contains long lived MS stars in the regime $3900~\text{K}<T_\text{eff}<5300~\text{K}$, where the MS lifetime is comparable to the age of the disc, i.e, stellar masses lower than the turn-off mass. For this purpose, we define the fraction $\delta$DM$_{\mathrm{DR2}}/\sigma_{1}$, which compares the uncertainty in distance modulus according to the Gaia DR2 parallax to the uncertainty in our model. In particular $\sigma_{1}$ is the result of our model with the best-fit parameters given in Table~\ref{table1} and Fig.~\ref{fig:corner} and indicates the precision we achieve in distance modulus. The uncertainty in distance modulus from error propagation of the Gaia DR2 parallax is 
\begin{equation}
\delta\text{DM}_{\mathrm{DR2}} = \frac{5}{\ln 10}\frac{\delta\varpi(G)_{\mathrm{DR2}}}{\varpi(\mathrm{DM_{phot}})},
\end{equation}

where $\delta\varpi(G)_{\mathrm{DR2}}$ is the expected parallax uncertainty in DR2. We estimate this by using the projected end of mission uncertainty as a function of G magnitude \citep[][their Fig.~10]{deBruijne}, as $\sqrt{3}\delta\varpi$. This takes into account the fact that we will have roughly 1/3 of the data after DR2. $\varpi(\mathrm{DM_{phot}})$ is the parallax that corresponds to the most likely photometric distance modulus to that star using our best fit model parameters given in Table~\ref{table1}. 

\begin{figure}
\centering
	\includegraphics[width=\columnwidth]{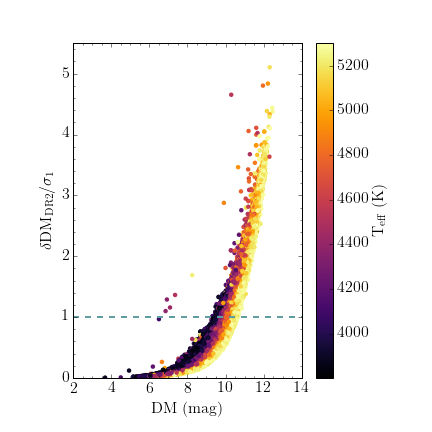}
    \caption{Distance modulus for stars in the regime $3800~\text{K} < T_\text{eff} < 5300~\text{K}$ in the L$\otimes$G sample compared to $\delta$DM$_{\mathrm{DR2}}/\sigma$, color coded by effective temperature. $\delta$DM$_{\mathrm{DR2}}/\sigma$ corresponds to the expected parallax uncertainty for DR2 divided by the parallax that corresponds to the most likely photometric distance to the star. Stars located above the dashed line are the stars for which our model performs better than DR2. We see a clear correlation with temperature.} 
    \label{fig:delta_DM_exp}
\end{figure}
We illustrate the results of this comparison in Fig~\ref{fig:delta_DM_exp}, where stars that lie above the dashed line with $\delta$DM$_{\mathrm{DR2}}/\sigma_{1} > 1$ are stars for which our model would perform better than Gaia DR2. We also observe that these results correlate with the effective temperature, showing that for warmer MS stars in our sample the fraction of stars that still need spectrophotometric distances is quite large. For intrinsically low-luminous (cool) stars the advantage of Gaia will be greatest. 
Faint stars in this survey will tend to be nearby which translates into larger parallaxes with low uncertainties. On the other hand, luminous stars will be observed up to larger distances, therefore at smaller parallaxes and with large uncertainties.

\section{Summary}
\label{sec:summary}

We have presented a method to calculate spectrophotometric distances for main sequence stars in the Milky Way, building a model from parallax and spectroscopic information. This model explicitly accounts for the parallax uncertainties in its construction, and accounts for the common binarity of near equal mass binaries among main sequence stars. Specifically, we construct a model for the mean absolute magnitude in the K band, $\overbar{M_K}$(T$_{\rm eff}$, $\log g$, [Fe/H]) from 4,000 MS stars with small ($\sigma_\varpi<$ 0.1) errors in parallax, in LAMOST $\times$ TGAS.

We then apply this model to LAMOST stars with very imprecise, or even no parallax information, obtaining good spectrophotometric distances for 150,000 MS stars in this sample. This work explicitly takes into account possible binarity of stars, which has not been explicitly considered in analogous models. Ignoring binaries could result in biased distances. 

We have built a model for the mean absolute magnitude of main sequence stars, which mostly 
draws on on parallax information whenever parallaxes are very informative ($\delta\varpi< 10\%$);
for increasingly poorer parallax estimates, this model gradually draws on the spectrophotometric information to estimate the distance modulus. We have shown that even 
in the regime of uninformative or missing parallaxes the model performs well, exploiting the information in the spectra: we obtain a value for the intrinsic dispersion in the absolute magnitude of single stars $
\sigma=$ 0.12 mag, which gives precisions in distance of $\sim$ 6\% for the fainter and more distant MS stars among current spectroscopic surveys.
We show that if we compare the distance moduli for the 150,000 stars in LAMOST $\times$ Gaia sample to the expected parallax uncertainties at the end of the Gaia mission presented in \citet{deBruijne} spectrophotometric distances are still needed. Especially for more luminous (and more distant) MS stars. 

As an application of precise distances we showed that they greatly improve the precision of orbital action estimates, as distance uncertainties dominate the orbit uncertainties with proper motions from Gaia.

\section*{Acknowledgements}
The thank the anonymous referee for suggestions that significantly improved a first version of this paper. J.C warmly thanks Sebasti\'an Bustamente for his help with the voroni plotting. This project was developed in part at the 2017 Heidelberg Gaia Sprint, hosted by the Max-Planck-Institut fur Astronomie, Heidelberg. J.C. acknowledges support from the SFB 881 program (A3) and the International Max Planck Research School for Astronomy and Cosmic Physics at Heidelberg University (IMPRS-HD). H.W.R. and W.H.T. received support from the European Research
Council under the European Union's Seventh Framework
Programme (FP 7) ERC Grant Agreement n. [321035]. This work has made use of data from the European
Space Agency (ESA) mission Gaia (http://www.cosmos.esa.int/gaia),
processed by the Gaia Data Processing and Analysis Consortium (DPAC,
http://www.cosmos.esa.int/web/gaia/dpac/consortium). Funding for the DPAC
has been provided by national institutions, in particular the institutions participating
in the Gaia Multilateral Agreement. 


\bibliographystyle{mnras}
\bibliography{example} 







\bsp	
\label{lastpage}
\end{document}